\documentstyle[epsfig]{mn}

\title[Gravitomagnetic corrections for spiral galaxy models]{Gravitomagnetic corrections to the lensing deflection angle for spiral galaxy models}
\author[S. Capozziello et al.]{S. Capozziello$^1$,
        V. F. Cardone$^1$,
        V. Re$^1$ and
        M. Sereno$^2$ \\
        $^1$Dipartimento di Fisica ``E.R. Caianiello'', Universit\`a di Salerno and INFN, Sezione di Napoli, Gruppo Collegato di Salerno, \\ 
Via S. Allende, 84081 - Baronissi (Salerno), Italy \\
	$^2$Dipartimento di Scienze Fisiche, Universit\`a degli Studi di Napoli ``Federico II'' and INFN, Sezione di Napoli, \\
Complesso Universitario di Monte S. Angelo, Via Cinthia - 80126 Napoli, Italy}

\date{Accepted xxx.
      Received yyy
      in original form zzz}

\begin{document}

\maketitle

\begin{abstract}

We investigate the effects of the gravitomagnetic corrections to the usual gravitational lens quantities for a specific lensing mass distribution modelled after spiral galaxies. An exponential disk is embedded into two different spherical halo models where disk and haloes parameters are fixed according to the observed mass\,to\,light ratios, galaxy magnitudes and rotation curves. The general expressions for the lensing deflection angle are given also taking into account the orientation of the galaxy disk plane with respect to the lens plane. It is found that the gravitomagnetic term changes the deflection angle by a typical amount of the order of ten microarcseconds.

\end{abstract}


\section{Introduction}

Gravitational lensing phenomena are a powerful tool to investigate the curvature of space\,-\,time nearby the source of gravitational field. The mathematical formulation of lensing is well developed. The weak field approximation may be used whenever the distance of the light ray to the lens is much larger than its Schwarzschild radius (e.g., $\stackrel{>}{\sim} 10$ times). This is the situation in several gravitational systems considered in astrophysical applications. When this is not the case, the mathematical formalism becomes much more complicated involving transcendental lens equations which cannot be easily solved (Bozza et al. 2001). Furthermore, interesting contributions to the lensing quantities, in the weak field approximation, could come out considering higher\,order corrections to the lens potential as the gravitomagnetic ones. Generally, such contributions are discarded since are considered too small \cite{SEF}. In Capozziello et al. (1999), these higher order corrections have been estimated for  point\,-\,like deflectors showing that they could give rise to non negligible effects. A further generalization was obtained by one of us (Sereno 2002). By using Fermat's principle and the standard assumptions of gravitational lensing, the gravitomagnetic corrections to the time delay function and the deflection angle for a geometrically thin lens were derived. The effects of the lens angular momentum on the propagation of light rays has been considered in literature using different approaches. For instance, Dymnikova (1986) determined the time delay of a signal due to the graviational field of a rotating body integrating the null geodesics of the Kerr metric. The effects of the rotation of the deflector have been evaluated to the lowest order by Glicenstein (1999) who used an argument based on Fermat's principle applied to the Lense\,-\,Thirring metric. These results are limited to pointlike deflectors and are difficult to generalize to more complex geometries. Asada \& Kasai (2000) used a multipole expansion of the deflecting potential to investigate the case of extended lenses. On the other hand, Sereno \& Cardone (2002) applied the formalism\footnote{This approach has been now further generalized to metric theories of gravity \cite{S03} to include also post\,-\,post\,-\,newtonian corrections in the computation of the deflection angle. However, in this paper, we will not consider these deviations from the standard general relativity theory.} developed in Sereno (2002) to the case of spherically symmetric mass distribution with angular momentum in the case of a rigid body rotation. 

The aim of this paper is to take into account the contribution of gravitomagnetic corrections in the case of spiral galaxies modelled according to two realistic mass profiles. We study how this contribution affects the lensing deflection angle and estimate its detectable range. This approach reveals particularly interesting, especially if related to the capabilities of the present 10\,m class telescopes (like VLT, Keck, Subaru), or those ones which are going to be available in the next future (like the 100\,m OWL telescope) and the space based interferometric mission, such as {\it SIM} and {\it GAIA}. The last decade has seen a notably effort in applying gravitational lensing as a tool to investigate the dynamical structure and morphology of galaxies. In particular, a main problem in astrophysics concerns the nature and the distribution of dark matter in the galactic components, that is the halo, the disk or the bulge. Paczy\'nski (1986) realized that this problem could be faced by gravitational microlensing. His predictions have been actually confirmed by the detection of several microlensing events towards the Magellanic Clouds \cite{Letal00,Aetal00}, the bulge of our Galaxy \cite{ogle} and the Andromeda Galaxy \cite{Point,Seb}. These can be considered as an evidence of the presence of dark (likely baryonic) components in the form of massive astrophysical compact halo objects (MACHOs). Anyway, MACHOs are not the only candidates for baryonic dark matter in the halo. It has been proposed, in fact, that an appreciable fraction could be in the form of diffuse objects as self gravitating gas clouds, with masses of the order of $10^{-3} \ M_{\odot}$ and radii $R \approx$\,10 AU \cite{PCM94,KBS02}. The presence of a dark halo is implied by the flat rotation curves in spiral galaxies \cite{SR01}. It has been suggested that the rotation curve of such galaxies follows the so\,-\,called {\it universal rotation curve} that can be characterized by a single free parameter, namely the total luminosity (or mass) of the disk \cite{PSS}. The profile of these curves is given by the sum of an exponential disk and a spherical halo  with a flat density core. However, this does not tell us what is the dark halo mass density, but it allows to recover some useful scaling relations to express the halo parameters as functions of those of the disk.
\\

This paper is organized as follows. In Sect.\,2, we describe the models of spiral galaxies which we consider and the scaling laws that we will use to determine their parameters. In Sect.\,3, we derive general expressions for the deflection angle of a spiral galaxy taking into account the gravitomagnetic corrections and the orientation of the disk plane with respect to the lens plane. Results are discussed in Sect.\,4, while conclusions are presented in Sect.\,5.

\section{Modelling spiral galaxies}

Following the standard approach, we will consider spiral galaxies as built by two main components\,: a disk and a dark halo. The disk, essentially, accounts for the optical properties  of the galaxy. Disk parameters may  be estimated by fitting the data taken in different photometric bands for a given model of the surface brightness distribution. It is usual to describe the disk surface density with the exponential profile\footnote{We assume the disk is idealised as being infinitely thin. This is not a serious shortcoming of our analysis since the finite thickness of the disk does not change the main results.} \cite{freeman}\,:

\begin{equation}
\Sigma(R) = \Sigma_{0} {\rm e}^{-R/R_d} \ .
\label{eq: diskphot}
\end{equation}
Two parameters are needed to  characterize the model\,: the central surface density $\Sigma_{0}$ and the scale length radius $R_d$. The latter is given by fitting the optical data, while $\Sigma_0$ is related to the observable central surface brightness $\mu_0 \equiv -2.5 \log{I_0}$ as\,:

\begin{equation}
\Sigma_0 = \Upsilon \ I_0 \ ,
\label{eq: mlratio}
\end{equation}
being $\Upsilon$ the disk mass\,-\,to\,-\,light ratio (which can also depend on the radius). Equation (\ref{eq: diskphot}) does not tell anything about the vertical density distribution, but this further quantity is not needed for our aims. The total disk mass is $M_{d} = 2 \pi \ \Sigma_{0} \ R_{d}^{2}$. Another important quantity is the rotation curve, i.e. the circular velocity of stars in the disk potential as function of the radius $R$ in the disk plane. For an exponential thin disk, it is \cite{freeman,BT87}\,:

\begin{equation}
v_{c, disk}(R) = \sqrt{\frac{2 G M_d}{R_d} y^2 \left [ I_0(y) K_0(y) - I_1(y) K_1(y) \right ]}
\label{eq: diskrot}
\end{equation}
with $y \equiv R/2R_d$; $I_n(y), K_n(y)$ are Bessel's functions of, respectively, first and second type of order $n$. The rotation curve increases until it reaches a maximum value and then decreases with a keplerian trend.

It is well known, however, that observed rotation curves of spiral galaxies remains flat well beyond the end of visible disk (see, e.g., Sofue \& Rubin 2001). Even if other theoretical solutions have been suggested (see, e.g., Milgrom 1983), the flatness of rotation curves may be easily explained by introducing a second non\,-\,visible component in spiral galaxies models\,: the dark halo. Deriving the shape and mass distribution of such a component from the analysis of the rotation curves is a highly degenerate problem since there are several halo models which fit equally well the observed rotation curve of a given spiral galaxy. Some hints on the halo structure come from N\,-\,body simulations of galaxy formation in hierarchical $\Lambda$CDM scenarios. However the results are still not unique due to problems of resolution and differences in the input basic properties of dark matter particles.

Among the many proposed distributions, we will consider here the so\,-\,called {\it Burkert\,-\,Borriello\,-\,Salucci model} (hereafter BBS; Burkert 1995; Borriello \& Salucci 2001; Salucci \& Borriello 2002), proposed to empirically account for the rotation curves of four dark matter dominated galaxies and successfully applied to a sample of 17 dwarf and LSB galaxies \cite{Ketal98}. Its density profile is\,:

\begin{equation}
\rho_{BBS}(r) = \frac{\rho_0 \ r_{0}^{3}}{(r + r_0) (r^2 + r_{0}^{2})} \ ,
\label{eq: bbsrho}
\end{equation}
where $\rho_0$ and $r_0$ are free parameters which represent the central dark matter density and the scale radius.

Since we adopt spherical symmetry, the halo contribution to the circular velocity is\,:

\begin{equation}
v_{c, BBS}(r) = \sqrt{\frac{G M_{BBS}(r)}{r}} \ .
\label{eq: bbsrot}
\end{equation}
being $M_{BBS}(r)$ the mass of the BBS halo enclosed within $r$, which is given by\,:

\begin{displaymath}
M_{BBS}(r) = 6.4 \ \rho_0 \ r_{0}^{3}\left \{ \ln{\left ( 1 + \frac{r}{r_0}
\right )} - \tan^{-1} {\left ( \frac{r}{r_0} \right )} +
\right .
\end{displaymath}
\begin{equation}
\left .
\ \ \ \ \ \ \ \ \ \ \ \ \ \ \ \ \ \ \ \ \ \ \ \ \ \ \ \ \ \ \ 
\frac{1}{2} \ln{\left [ 1 + \left ( \frac{r}{r_0} \right )^2 \right ]} \right \} \ .
\label{eq: bbsmass}
\end{equation}
The mass distribution diverges so that we need to introduce a cut\,-\,off radius in order to get a finite total mass. To this aim, we will truncate the halo at the virial radius $r_{vir}$ defined as the radius within which the mean density is $\delta_{vir} \ \rho_{crit}$, with $\delta_{vir}$ depending on the chosen cosmological model and $\rho_{crit} = 3H_0^2/8 \pi G$ the critical density of the universe. The total mass $M_{vir}$ is thus obtained by putting $r = r_{vir}$ in Eq.(\ref{eq: bbsmass}). To estimate $r_{vir}$, we have to fix the value of $\delta_{vir}$ choosing a background cosmological model. We adopt the currently popular $\Lambda$CDM model with $(\Omega_m, \Omega_{\Lambda}) = (0.3, 0.7)$ when $\delta_{vir} = 337$.

In principle, there is no obvious reason why disk and halo parameters should be correlated. Actually, this turns out to be true\,: a clear correlation exists. By considering a sample of 1100 synthetic rotation curves built from 15000 measurements, Persic et al. (1996a,b) have demonstrated the existence of the so\,-\,called {\it universal rotation curve} (hereafter URC), a function which describes (within the measurement errors) the rotation curve of all spiral galaxies, modulo a scaling factor related to the disk total luminosity. From the very existence of the URC, it is possible to infer some useful scaling relations among halo and disk parameters. Briefly, one decomposes the URC as\,:

\begin{equation}
v_{URC}^{2} = v_{c, disk}^{2} +  v_{c, halo}^{2}
\label{eq: vurc}
\end{equation}
with $v_{c, disk}$ as in Eq.(\ref{eq: diskrot}) and $v_{c, halo}$ determined by the dark halo model. The universality of the URC does imply that disk and halo parameters should be related by some scaling relations in order to reproduce the URC itself. Modelling the halo by the BBS profile, Salucci \& Burkert (2000) have found the following scaling relations among the halo parameters and the total disk mass (see also Salucci \& Borriello 2002)\,:

\begin{equation}
\log{\rho_0} = -23.0 -0.077 \ \log{M_d} - 9.98 {\times} 10^{-6} \ M_{d}^{0.43} \ ,
\label{eq: rhomd}
\end{equation}

\begin{equation}
\log{r_0} = 9.10 + 0.28 \ \log{\rho_0} - 3.49 {\times} 10^{10} \ \rho_{0}^{0.43}\,, \label{eq: rhorz}
\end{equation}
with $\rho_{0}$ in g\,cm$^{-3}$, $M_d$ in units of the solar mass $M_{\odot}$ and $r_0$ in kpc. Adding to Eqs.(\ref{eq: rhomd}, \ref{eq: rhorz}) a third relation \cite{PSS,SB02}\,:

\begin{equation}
\log{R_d} = 4.96 - 1.17 \ \log{M_d} + 0.070 \ (\log{M_d})^2 \ ,
\label{eq: mdrd}
\end{equation}
it is  possible to estimate all the galaxy parameters from the knowledge of the disk total mass. In principle, it should  be possible to use the disk scale length $R_d$ as an order parameter since it is directly measurable. However, it is worth to remember that it may vary with the photometric band chosen to fit the disk model. On the other hand, the disk total mass $M_d$ may be indirectly obtained by measuring the central surface brightness $\mu_0$ and using Eq.(\ref{eq: mlratio}) since $\Upsilon$ is independently estimated. It is worthwhile to stress that there are different possibilities to get an estimate of $\Upsilon$. We only remember here the correlations between $\Upsilon$ and the color $B - R$ \cite{BdB01} or between $\Upsilon$ and $R_d$ \cite{G02}.

The BBS model is not the only density profile able to reproduce the observed URC. Actually, in their original paper, Persic et al. (1996a) proposed a different approach to the URC by modeling directly the halo contribution to the rotation curve. They used\,:

\begin{equation}
v_{c, halo}^{2}(x) = v_{opt}^{2} (1 - \beta_h) (1 + a^2) \frac{x^2}{x^2 + a^2} \ ,
\label{eq: vhalopss}
\end{equation}
with $v_{opt}$ the circular velocity at the optical radius $R_{opt} = 3.2 R_d$, $x = r/R_{opt}$, $a$ the halo core radius in units of $R_{opt}$ and\,:

\begin{equation}
\beta_h = \frac{1.1 \ G \ M_d}{v_{opt}^{2} \ R_d} \ .
\label{eq: beta}
\end{equation}
If we assume spherical symmetry, the mass distribution corresponding to the halo rotation curve in Eq.(\ref{eq: vhalopss}) is\,:

\begin{equation}
M_{PSS}(x) = \frac{R_{opt} v_{opt}^{2}}{G} (1 - \beta_h) (1 + a^2) \frac{x^3}{x^2 + a^2} \ ,
\label{eq: pssmass}
\end{equation}
which is once again divergent. As before, we will truncate the halo at the virial radius $r_{vir}$. The mass inside $r_{vir}$ is thus the total mass of the halo. We will refer, in the following, to this model as the {\it PSS model}. The density profile turns out to be\,:

\begin{equation}
\rho_{PSS}(r) = \frac{v_{opt}^2}{2 \pi G R_{opt}^2} (1 - \beta_h) (1 + a^2) \frac{a^2}{x (a^2 + x^2)^2} \ .
\label{eq: rhopss}
\end{equation}
The URC is then obtained by using the following set of scaling relations (Persic et al., 1996a,b)\,:

\begin{displaymath}
\Sigma_0 = \frac{M_d}{2 \pi R_{d}^2} \ = 7.8 {\times} 10^8 \ \left ( 1 + 0.6 \ \log{\frac{L_I}{L_{\star}}} \right )^2
\end{displaymath}
\begin{equation}
\ \ \ \ \ \ \ \ \ \ \ \ \ \ \ \ \ \ 
= 7.8 {\times} 10^8 \ [ 1 - 0.4 \ (M_I + 21.9)]^2 \ ,
\label{eq: sigmazeropss}
\end{equation}

\begin{equation}
\beta_h = 0.72 + 0.44 \ \log{\frac{L_I}{L_{\star}}} = 0.72 - 0.176 \ (M_I + 21.9) \ ,
\label{eq: betapss}
\end{equation}

\begin{equation}
a = 1.5 \ \left ( \frac{L_I}{L_{\star}} \right )^{0.2} = 1.5 {\times} 10^{-0.016 \ (M_I + 21.9)}
\label{eq: apss}
\end{equation}
where $L_I$ is the total luminosity (in solar units) in the $I$ band, $M_I$ the absolute total magnitude and $-2.5 \log{L_{\star}} = -21.9$. We will use $M_I$ as an order parameter since it can be measured from observations; for each value of $M_I$, we first use Eqs.(\ref{eq: mdrd}) and (\ref{eq: sigmazeropss}) to get the disk parameters. Then Eqs.(\ref{eq: beta}), (\ref{eq: betapss}) and (\ref{eq: apss}) will allow us to estimate the halo parameters. So the spiral galaxy model is fully characterized.

The two halo models described here will be used in the following section to estimate whether the gravitomagnetic corrections may be directly observed in realistic cases. The scaling relations quoted above will help us to estimate the effects as a function of only one ordering parameter: the disk total mass $M_d$ in  the BBS case and the disk absolute total magnitude $M_I$ in the PSS model.

\section{Lensing deflection angle}

Following Sereno (2002), the deflection angle to the order $c^{-3}$ reads\,:

\begin{equation}
\vec{\alpha}( \vec{\xi} ) = \frac{4 G}{c^2} \int_{\Re ^2}
{d^2\xi' \Sigma( \vec{\xi}' ) \ \left [ 1 - 2 \frac{\bar{v}_l( \vec{\xi}' )}{c} \right ] \frac{\vec{\xi} - \vec{\xi}'}{| \vec{\xi} - \vec{\xi}' |^2}}
\label{eq: alphagrav}
\end{equation}
being $\bar{v_l}( \vec{\xi}' )$ the weighted average, along the line of sight, of the component of the velocity orthogonal to the lens plane, i.e.\,: 

\begin{equation}
\bar{v}_l( \vec{\xi} ) \equiv \frac{\int{(\vec{v}( \vec{\xi}, l ) \cdot \vec{e}_{in}) \ \rho( \vec{\xi}, l ) dl}}
{\Sigma( \vec{\xi} )} \ ,
\label{eq: defvmedio}
\end{equation}
and\,:

\begin{equation}
\Sigma( \vec{\xi} ) = \int_{-\infty}^{+\infty}{\rho( \vec{\xi}, l ) dl} \ .
\label{eq: defsigma}
\end{equation}
In Eqs. (\ref{eq: alphagrav}), (\ref{eq: defvmedio}) and (\ref{eq: defsigma}), we are using a coordinate system centred on the galaxy lens with the axes $(\xi_1, \xi_2)$ in the lens plane which is ortogonal to the direction of the incoming light ray $\vec{e}_{in}$ (coincident with the line of sight), while $l$ is the component along the line of sight. Eqs.(\ref{eq: alphagrav}) and (\ref{eq: defvmedio}) show that, in the thin lens approximation, motions in the plane of the lens does not give any contribution to the deflection angle. 

Due to tidal interactions with other galaxies during its formation process, a spiral galaxy has a non\,-\,zero angular momentum which gives rise to corrections to the lensing deflection angle. The effects of the lens angular momentum on the propagation of light rays has been considered by Sereno \& Cardone (2002). They applied the formalism developed in Sereno (2002) to the case of a rotating spherically symmetric lens in the approximation of rigid rotation, i.e. assuming that the angular velocity does not depend on coordinates. While useful to estimate the gravitomagnetic corrections for lensing by stars, the results in Sereno \& Cardone (2002) have to be generalized to the case of non costant angular velocity to be applied to spiral galaxies.  

To this aim, let us consider a spherically symmetric lens that rotates about an arbitrary axis, $\hat{\eta}$, passing through its centre (i.e. a main axis of inertia). To specify the orientation of the rotation axis, we need two Euler's angles\,: $\beta$ is the angle between the line of sight $\hat{l}$ and the line of nodes defined at the intersection of the $\hat{l \ \xi_1}$ plane and the equatorial plane (i.e., the plane orthogonal to the rotation axis and containing the lens centre); $\gamma$ is the angle between $\hat{\eta}$ and the $\xi_2$\,-\,axis. Using the axial symmetry about the rotation axis, one easily gets \cite{SCnoi02}\,:

\begin{equation}
{\bf v} \cdot {\bf e}_{in}(\xi_1, \xi_2, l) =
- \omega(R) \left [ \xi_1 \cos{\gamma} + \xi_2 \cos{\beta} \sin{\gamma} \right ] 
\label{eq: vcomp}
\end{equation}
where $\omega (R)$ is the modulus of the angular velocity at a distance $R \equiv (R_1^2+R_2^2)^{1/2}$ from the rotation axis, while $\hat{R}_1$ (that, given the spherical symmetry of the system, can be taken along the line of nodes) and $\hat{R}_2$ are the axes on the equatorial plane. It is\,:

\begin{equation}
R_1 = l \cos{\beta} + \xi_1 \sin{\beta} \ ,
\label{eq: runo}
\end{equation}

\begin{equation}
R_2 =- l \sin{\beta} \cos{\gamma} + \xi_1 \cos{\beta} \cos{\gamma} + \xi_2 \sin{\gamma} \  .
\label{eq: rdue}
\end{equation}
For a self\,-\,gravitating spherically symmetric system, both the energy and the angular momentum are integrals of motion and the ordered motion of the stars is on the plane orthogonal to the rotation axis. Thus the angular velocity reads\,:

\begin{equation}
\omega(R) = \frac{v_c(R)}{R} = 
\frac{\sqrt{v_{c, disk}^2(R) + v_{c, halo}^2(R)}}{R} \ .
\label{eq: vcomega}
\end{equation}
Note that the circular velocity entering Eq.(\ref{eq: vcomega}) is the total one, i.e. the sum of the contributions due to both the luminous disk and the dark halo. Inserting Eqs.(\ref{eq: vcomega}) into Eq.(\ref{eq: vcomp}) and the result into Eq.(\ref{eq: alphagrav}), it is straightforward to get the following general expressions for the two components of the deflection angle\,:

\begin{equation}
\alpha_1(\xi, \theta) = \alpha_1^{(0)} + \frac{8 G}{c^3} 
\left ( I_{1a} \cos{\gamma} + I_{1b} \cos{\beta} \sin{\gamma} \right ) \ ,
\label{eq: alphaoneaxi}
\end{equation}

\begin{equation}
\alpha_2(\xi, \theta) = \alpha_2^{(0)} + \frac{8 G}{c^3} 
\left ( I_{2a} \cos{\gamma} + I_{2b} \cos{\beta} \sin{\gamma} \right ) \ ,
\label{eq: alphatwoaxi}
\end{equation}
where we have introduced polar coordinates $(\xi, \theta)$ in the lens plane. Here, $\alpha_i^{(0)}$ are the two components of the classical deflection angle which, for axisymmetric systems, turn out to be\,:

\begin{equation}
\alpha_1^{(0)} = \frac{4 \pi G}{c^2 \xi} \cos{\theta} \int_0^{\xi}{\Sigma(\xi') \xi' d\xi'} \ ,
\label{eq: alphaclassone}
\end{equation}

\begin{equation}
\alpha_1^{(0)} = \frac{4 \pi G}{c^2 \xi} \sin{\theta} \int_0^{\xi}{\Sigma(\xi') \xi' d\xi'} \ .
\label{eq: alphaclasstwo}
\end{equation}
In Eq.(\ref{eq: alphaoneaxi}) it is\,:
\begin{displaymath}
I_{1a}(\xi, \theta) \equiv \int_{0}^{\infty}{\Sigma(\xi') \xi'^2 d\xi'} \ \times
\end{displaymath}
\begin{equation}
\ \ \ \ \ \ \ \ \ \ \ \ 
\int_{0}^{2 \pi}{\frac{\left ( \xi \cos{\theta} - \xi' \cos{\theta'} \right ) \cos{\theta'}}
{\xi^2 + \xi'^2 - 2 \xi \xi' \cos{(\theta - \theta')}} \ \langle \omega \rangle (\xi', \theta') d\theta'} \ ,
\label{eq: defiunoa}
\end{equation}

\begin{displaymath}
I_{1b}(\xi, \theta) \equiv \int_{0}^{\infty}{\Sigma(\xi') \xi'^2 d\xi'} \ \times
\end{displaymath}
\begin{equation}
\ \ \ \ \ \ \ \ \ \ \ \ 
\int_{0}^{2 \pi}{\frac{\left ( \xi \cos{\theta} - \xi' \cos{\theta'} \right ) \sin{\theta'}}
{\xi^2 + \xi'^2 - 2 \xi \xi' \cos{(\theta - \theta')}} \ \langle \omega \rangle (\xi', \theta') d\theta'} \ ,
\label{eq: defiunob}
\end{equation}
while, in Eq.(\ref{eq: alphatwoaxi}), $I_{2a}$ ($I_{2b}$) is defined as $I_{1a}$ ($I_{1b}$) by substituting the $\cos{\theta'}$ with $\sin{\theta'}$ in the brackets on the right hand side. The average angular velocity\footnote{Note that, if we assume solid body rotation, $\omega$ is constant and may be taken out from the integrals in Eqs.(\ref{eq: defiunoa}), (\ref{eq: defiunob}). In this case, Eqs.(\ref{eq: alphaoneaxi}) and (\ref{eq: alphatwoaxi}) reduce to Eqs.(8) and (9) of Sereno \& Cardone (2002).} has been defined as\,:

\begin{equation}
\langle \omega \rangle(\xi, \theta) = \frac{1}{\Sigma(\xi)} 
\int_{-\infty}^{\infty}{\frac{v_c(\xi, \theta, l) \rho(\xi, l)}{\sqrt{\xi^2 + l^2}} dl}
\label{eq: omegamedio}
\end{equation}
having used the axial symmetry of the halo models described in the previous section. Note that $v_c$ depends on $(\xi, \theta, l)$ through $R$ because of Eqs.(\ref{eq: runo}) and (\ref{eq: rdue}). 

The gravitomanetic corrections we are investigating are the terms $I_{1a,b}$ and $I_{2a,b}$ that are of order $v_c/c$. It is now straightforward to compute the deflection angle for a spiral galaxy considered as a two component system, i.e. an exponential disk and a dark halo (modelled with the BBS or the PSS profile). The results for the two different modelling are presented and discussed in the next section.

\section{Results}

To study the detectability of the gravitomagnetic corrections, we have to determine the images positions in order to investigate whether they are significantly modified by these higher order terms. This could be a difficult task since the lens equations are highly nonlinear and the results depend also on the source position. However, the images positions are related to the deflection angle so that, if this latter is not modified significantly by gravitomagnetic corrections, we have no detectable effects on the images angular separation which is the easiest quantity to measure. We can thus investigate whether the higher order terms introduce appreciable corrections to the total deflection angle. To this aim, we introduce the quantities\,:

\begin{figure}
\centering
\resizebox{8.5cm}{!}{\includegraphics{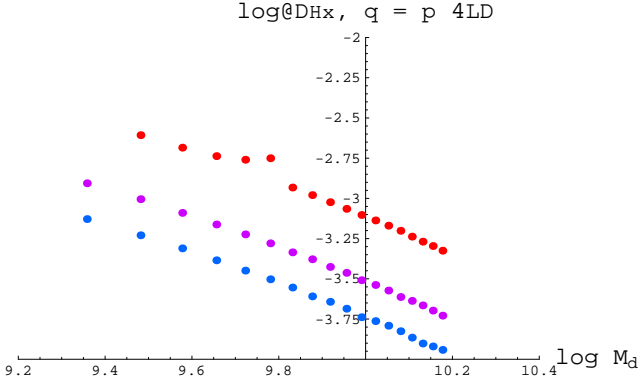}}
\hfill
\caption{$\log{|\Delta|}$ vs $\log{M_d}$ for the BBS model. The curves are referred to $\xi/r_{vir} = 0.5, 1.0, 1.5$ from top to bottom, while $\theta$ is fixed to $\pi/4$.}
\label{fig: deltabbs}
\end{figure}

\begin{equation}
\Delta_i(\xi, \theta; M_d) = 100 \ \times \ 
\frac{\alpha_i(\xi, \theta) - \alpha_i^{(0)}(\xi, \theta)}{\alpha_i(\xi, \theta)} 
\ \ , \ \ i = 1,2  \ ,
\label{eq: deltai}
\end{equation}

\begin{equation}
\Delta(\xi, \theta; M_d) = 100 \ \times \ 
\frac{\alpha(\xi, \theta) - \alpha^{(0)}(\xi, \theta)}{\alpha(\xi, \theta)}
\label{eq: deltamod}
\end{equation}
being $\alpha = (\alpha_1^2 + \alpha_2^2)^{1/2}$ the modulus of the deflection angle. Note that this quantity is now function of $(\xi, \theta)$ and not only of $\xi$ since the gravitomagnetic corrections break down the radial symmetry of the deflection angle \cite{S02,SCnoi02}. In Eqs.(\ref{eq: deltai}) and (\ref{eq: deltamod}), we consider only the dependence on the disk mass $M_d$ since the scaling relations described in Sect.\,2 allow us to fully characterize the model (i.e. disk and halo parameters) as function of this latter quantity. For the PSS model, the ordering parameter is the disk magnitude, but there is a one\,-\,to\,-\,one correspondence between $M_I$ and $M_d$ so that we will use $M_d$ to compare the results for the BBS and the PSS models. To estimate $\Delta_i$ and $\Delta$ we have to fix the values of $(\xi, \theta)$. Since the deflection angle is roughly proportional to the projected mass inside $\xi$, we expect the corrections to be most important for larger values of $\xi$. On the other hand, the deflection angle also depends on the angular coordinate $\theta$. We have decided to compute $\Delta_i$ and $\Delta$ for three values of $\xi$, namely $\xi/r_{vir} = 0.5, 1.0, 1.5$, while we fix $\theta = \pi/4$ to speed up the calculations\footnote{Note that, for this value of $\theta$, we get\,: $\Delta_1 = \Delta_2$ and $\Delta = \sqrt{2} \Delta_1$. We do not explicitely calculated $\Delta_i$ and $\Delta$ for other values of $\theta$ since it is reasonable to expect that $\Delta(\xi, \theta \ne \pi/4)$ is of the same order of magnitude as $\Delta(\xi, \theta = \pi/4)$ that is enough for our aims.}. Finally, we have also to choose the orientation of the disk plane with respect to the lens one. We arbitrarily fix $\beta = 0$ and $\gamma = \pi/4$. It is reasonable to expect that choosing different values of $(\beta, \gamma)$ does not change significantly our results. When computing $\Delta_i$ and $\Delta$, we have included the disk contribution to the classical deflection angle, but we have neglected the gravitomagnetic corrections for the disk. In fact, we have checked that the classical deflection angle of the disk is of the same order of magnitude as the gravitomagnetic corrections for the halo at the positions where we calculate $\Delta_i$ and $\Delta$. Hence, the gravitomagnetic corrections for the disk will be much smaller than those for the halo and can be safely neglected\footnote{Actually, this could not be true for high inclination angles\,: disks almost perpendicular to the plane of the sky may have such a high projected surface mass density that they play a dominant role in the appearance of lensed image configurations \cite{MFP97,BL98}. For these configurations, we should take into account also the disk contribution to the gravitomagnetic corrections. However, it is quite unlikely that this alters significantly our main results since deflection angles will be estimated at $\xi \ge 0.5 \ r_{vir}$ that is far from the disk edge since $r_{vir} >> R_d$. For these values of the impact parameter, the disk rotational velocity has almost vanished and hence the gravitomagnetic corrections due to the disk could be neglected.}. We stress, however, that the disk plays a key role in our analysis. First, it is the disk mass $M_d$ which, through the scaling relations discussed in Sect.\,2, determines the halo parameters and hence the gravitomagnetic corrections. Second, the disk potential contributes to the total circular velocity entering Eq.(\ref{eq: omegamedio}). It is thus important to include the disk in our analysis, although we neglect its contribution to the gravitomagnetic terms\footnote{This also saves computational time without introducing any significative systematic error.}.
\begin{figure}
\centering
\resizebox{8.5cm}{!}{\includegraphics{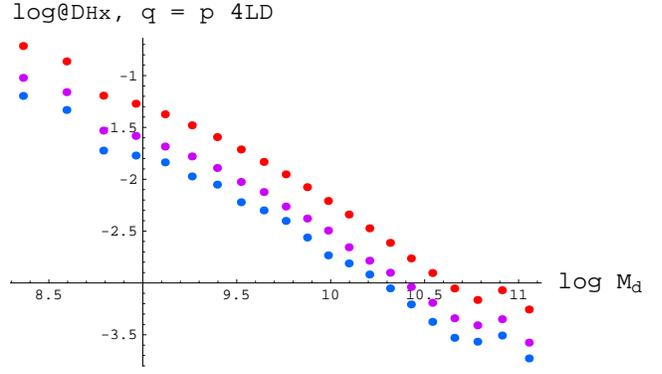}}
\hfill
\caption{Same as Fig.\,\ref{fig: deltabbs} but for the PSS model.}
\label{fig: deltapss}
\end{figure}
In Figs.\,\ref{fig: deltabbs} and\,\ref{fig: deltapss}, we show $\Delta$ as function of the disk mass $M_d$; a log\,-\,log plot is used to better illustrate the results\footnote{Note that $\Delta(\xi, \theta)$ is always negative for the cases reported in Figs.\ref{fig: deltabbs} and\,\ref{fig: deltapss} so that the plots report $\log{|\Delta|}$. However, this is not a general result depending on the particular choice of the orietantion angles $(\beta, \gamma)$.}. Despite the different halos and scaling laws used, the results for the two models are qualitatively similar. This is not an unexpected result. Eqs.(\ref{eq: alphaoneaxi})\,--\,(\ref{eq: omegamedio}) show that the gravitomagnetic corrections depend on the circular velocity $v_c(R)$ of the model averaged along the line of sight. On the other hand, the parameters of both the BBS and the PSS models have been fixed so that the two models reproduce the same URC. As a consequence, for a fixed value of $M_d$, the functional form of $v_c(R)$ and thus of $\langle \omega \rangle(\xi, \theta)$ is the same and hence $\Delta$ has the same trend for the two different halo models. The magnitude of the percentual correction is however quite different, the PSS one being almost two orders of magnitude higher. This result reflects the difference between the BBS and the PSS model. Qualitatively, $\Delta \propto 1/\alpha \sim 1/\alpha^{(0)} \propto M^{-1}(\xi)$ and hence the higher is the mass enclosed within $\xi$, the lower is the percentage correction. Since, for a fixed $M_d$, the value of $M(\xi)$ for the BBS model is higher than that for the PSS halo due to the fact that the BBS mass distribution is more concentrated, it turns out that $\Delta$ is larger for the PSS model.

As a final remark, we note that the formulae in Sect.\,2 may be used, in principle, whatever is the astrophysical system considered provided that its surface mass distribution $\Sigma(\xi)$ is given and the system is spherically symmetric. A typical example is a cluster of galaxies which could be modelled using the NFW density profile \cite{NFW97} and opportune scaling relations. Since the gravitomagnetic corrections roughly scales as $v_c/c$, one could suppose that $\Delta$ should be higher for a cluster of galaxies rather than for a single spiral galaxy. A simple scaling argument may qualitatively show that this is not true at all. The ratio of th gravitomagnetic corrections is roughly\,:

\begin{displaymath}
\frac{\left [ \alpha - \alpha^{(0)} \right ]_{galaxy}}{\left [ \alpha - \alpha^{(0)} \right ]_{cluster}} 
\sim \frac{(v/c)_{galaxy}}{(v/c)_{cluster}} \sim 10^{-1} \ ,
\end{displaymath} 
while for the total deflection angle we get approximately\,:

\begin{displaymath}
\frac{\alpha_{galaxy}}{\alpha_{cluster}} \sim \frac{\alpha^{(0)}_{galaxy}}{\alpha^{(0)}_{cluster}} \sim
\frac{M_{galaxy}}{M_{cluster}} \sim 10^{-3} \ .
\end{displaymath}
Thus we obtain\,:

\begin{displaymath}
\frac{\Delta_{galaxy}}{\Delta_{cluster}} = 
\frac{\left [ \left ( \alpha - \alpha^{(0)} \right )/\alpha \right ]_{galaxy}}
{\left [ \left ( \alpha - \alpha^{(0)} \right )/\alpha \right ]_{cluster}} \sim 100
\end{displaymath}
so that the gravitomagnetic corrections for a cluster of galaxies are two orders of magnitude lower than for a spiral galaxy. Some attempts, using Eqs.(\ref{eq: alphagrav})\,--\,(\ref{eq: omegamedio}) and the NFW density profile, have strenghtened this qualitative result.

\section{Conclusions}

The detectability of the gravitomagnetic corrections we have evaluated is a hard task. Figs.\,1\,-\,2 show that, for the BBS model, $| \Delta |$ may be as high as $0.002\%$, while, for the PSS model, the corrections ranges up to $0.16\%$. These values refer to percentage corrections. It is thus useful to give some absolute estimates of the gravitomagnetic corrections. For the BBS model we get\,:

\begin{displaymath}
| \alpha - \alpha^0 | \sim 21 \div 857 \ \mu as \ \ \ \ {\rm for} \ \xi/r_{vir} = 0.5 \ ,
\end{displaymath}

\begin{displaymath}
| \alpha - \alpha^0 | \sim 5 \div 71 \ \mu as \ \ \ \ {\rm for} \ \xi/r_{vir} = 1.0 \ ,
\end{displaymath}

\begin{displaymath}
| \alpha - \alpha^0 | \sim 2 \div 7 \ \mu as \ \ \ \ {\rm for} \ \xi/r_{vir} = 1.5 \ ,
\end{displaymath}
while for the PSS model it is\,:

\begin{displaymath}
| \alpha - \alpha^0 | \sim 29 \div 350 \ \mu as \ \ \ \ {\rm for} \ \xi/r_{vir} = 0.5 \ ,
\end{displaymath}

\begin{displaymath}
| \alpha - \alpha^0 | \sim 8 \div 94 \ \mu as \ \ \ \ {\rm for} \ \xi/r_{vir} = 1.0 \ ,
\end{displaymath}

\begin{displaymath}
| \alpha - \alpha^0 | \sim 3 \div 43 \ \mu as \ \ \ \ {\rm for} \ \xi/r_{vir} = 1.5 \ ,
\end{displaymath}
having evaluated all these quantities for $\theta = \pi/4$ and $(\beta, \gamma) = (0, \pi/4)$. The images positions depend implicitely on the deflection angles so that it is difficult to estimate the variation of the images angular separations $\Delta \theta$ due to the higher order terms. If one assumes that the correction to $\alpha(\xi, \theta)$ leads to a similar variation of $\Delta \theta$, then the prospect to detect the effects of the gravitomagnetic terms could be considered quite good. Actually, the astrometric precision of the NASA {\it Space Interferometric Mission} (to be launched in 2009) is estimated to be $\sim 4 \ \mu as$, while the european {\it GAIA} satellite (scheduled for 2010) will be able to measure stars position with an accuracy of $\sim 1 \ \mu as$. Both these satellites will be thus able to detect tiny deviations in the images positions from the values predicted by the standard lensing theory. These instrumentations can thus lead these higher order terms in the realm of detectability. 

In addition to the rotational velocity, galaxies typically have also a peculiar line\,-\,of\,-\,sight velocity whose effect on the deflection angle is easy to estimate. We simply write the total deflection angle as the sum of the standard term, $\alpha^{(0)}$ and the gravitomagnetic correction, $\alpha^{GRM}$. This latter may then be splitted as $\alpha^{GRM} = \alpha^{GRM}_{rot} + \alpha^{GRM}_{pec}$, being $\alpha^{GRM}_{rot}$ the contribute of the lens angular momentum and $\alpha^{GRM}_{pec}$ the term due to the peculiar motions. Sereno (2002) has shown that\.:

\begin{displaymath}
\alpha^{GRM}_{pec} = - 2 \alpha^{(0)} \frac{v_{pec}}{c} \ .
\end{displaymath}
For a receding lens, $v_{pec} < 0$ so that the effect of the lens peculiar motion increases the gravitomagnetic correction thus strenghtening the possibility of detection. 

On the other hand, the gravitomagnetic corrections due to the dark halo angular momentum could also be lower than we have evaluated. A key ingredient in the estimate of the higher order terms through Eqs.(\ref{eq: defiunoa}), (\ref{eq: defiunob}) is the average angular velocity $\langle \omega \rangle$ that we have computed using Eqs. (\ref{eq: vcomega}) and (\ref{eq: omegamedio}). This approach implicitely assumes that the halo objects have only an ordered motion. While this is a correct assumption for the disk particles, the high value of the halo velocity dispersion suggests that random motions could play a significative role in determining the angular momentum $J$. As a test, one could evaluate the spin parameter $\lambda = J |E|^{1/2}/G M_{vir}^{5/2}$, with $E$ the total energy \cite{MMW98}, and compare it to the results of the simulations of galaxy formation (see, e.g., Vitvitska et al. 2002). We find that $\lambda$ is overestimated by a factor of order $\sim 7$ thus suggesting that $J$ is overestimated too. Should this come to be true, the values of the gravitomagnetic corrections we have reported should be lowered by a factor of order $\sim 1/7$. However, it is also possible that the angular momentum of the nowaday halo is higher than that of the protogalactic halo because of the exchange of angular momentum between the halo and the collapsed baryons. 

Even if we neglect this question, however, it is worth noting that the variations of the deflection angle due to the gravitomagnetic corrections are smaller or of the same order as the ones due to more classical effects. Uncertainties in the value of the model parameters may induce errors on the value of $\alpha^{(0)}$ and these latter could be so high that the gravitomagnetic corrections are completely masked. Even if the model parameters were known with high accuracy through independent techniques, an external perturbing shear (due to, e.g., the cluster of galaxies which the main lens belongs to or to large scale structure) could mimic the same effect as the higher order terms we have considered in this paper. On the other hand, this problem may also be reversed. There are many multiply imaged QSOs systems where there are no evidences of a cluster of galaxies, but nonetheless a small external shear is needed to fit the observed images configuration. Deviations from axisymmetry in the lens galaxy are usually claimed to be the origin of this perturbing shear, but there are no definitive evidences favouring this hypothesis. It should be interesting to reconsider these systems to see whether the gravitomagnetic corrections (whose effects could be of the same order as that of a small shear) could reconcile model and data without invoking this external perturbation. 

Our results are encouraging since they suggest that the gravitomagetic corrections may indeed be detectable. However, a further analysis is needed in order to investigate how the images position is modified and whether corrections emerge also for other lensing observables (such as the amplification). Actually, it is especially interesting to consider the effect of higher order terms on time delay $\Delta t$ between the images since this is one of the most promising direct technique for the measurement of the Hubble constant \cite{Ref64,Herqules}. The time delay $\Delta t$ roughly scales with the angular separation $\Delta \theta$ between the images (see, e.g., Schneider et al. 1992). Since the gravitomagnetic corrections alter the lensing deflection angle, these higher order terms could change $\Delta \theta$ and thus $\Delta t$ leading to a different estimation of $H_0$. Furthermore, the reconstruction of the lensing potential from images configuration can be affected by the gravitomagnetic corrections since the deflecting potential $\psi( \vec{\xi} )$, which is used to evaluate the quantity to be fitted, is modified \cite{S02}. Neglecting gravitomagnetic corrections thus introduces a systematic error in the determination of the Hubble constant. Investigating in detail this issue will be the topic of a forthcoming paper. 

\section*{acknowledgements}

One of the authors (SC) wishes to acknowledge Paolo Salucci and Luigi Danese for the useful discussions and suggestions on the topics of this paper. We also thank an anonymous referee whose constructive comments have helped us to significantly improve the paper.


\begin{thebibliography}{99}

\bibitem[\protect\citename{Alcock et al. }2001]{Aetal00}
Alcock, Ch. et al. 2001, ApJ, 542, 281

\bibitem[\protect\citename{Asada \& Kasai }2000]{AK00}
Asada, H., Kasai, M. 2000, Progr. Theor. Phys, 104, 95

\bibitem[\protect\citename{Auriere et al. }2001]{Point}
Auri\`{e}re, M. et al. 2001, ApJ, 553, L137

\bibitem[\protect\citename{Bartelmann \& Loeb }1998]{BL98}
Bartelmann, M., Loeb, A. 1998, ApJ, 503, 48

\bibitem[\protect\citename{Bell \& de Jong }2001]{BdB01}
Bell, E.F., de Jong, R.S. 2001, ApJ, 550, 212

\bibitem[\protect\citename{Binney \& Tremaine }1987]{BT87}
Binney, J., Tremaine, S. 1987, {\it Galactic dynamics}, Princeton University Press, Princeton.

\bibitem[\protect\citename{Borriello \& Salucci }2001]{BS01}
Borriello, A., Salucci, P. 2001, MNRAS, 323, 285

\bibitem[\protect\citename{Bozza et al. }2001]{bozza}
Bozza, V., Capozziello, S., Iovane, G., Scarpetta, G. 2001, Gen. Rel. Grav., 33, 1535

\bibitem[\protect\citename{Burkert }1995]{B95}
Burkert, A. 1995, ApJ, 447, L25

\bibitem[\protect\citename{Burkert \& Silk }1997]{BuS97}
Burkert, A., Silk, J. 1997, ApJ, 488, L55

\bibitem[\protect\citename{Calchi Novati et al. }2002]{Seb}
Calchi Novati, S. et al. 2002, A\&A, 381, 848

\bibitem[\protect\citename{Capozziello et al. }1999]{CLPS99}
Capozziello, S., Lambiase, G., Papini, G., Scarpetta, G. 1999, Phys. Lett. A, 254, 11.

\bibitem[\protect\citename{Cardone et al. }2002]{Herqules}
Cardone, V.F., Capozziello, S., Re, V., Piedipalumbo, E. 2002, A\&A, 382, 792

\bibitem[\protect\citename{Dymnikova }1986]{D86}
Dymnikova, I. 1986, in {\it Relativity in celestial mechanics and astrometry}, eds. J. Kovalevsky, A. Brumberg

\bibitem[\protect\citename{Freeman }1970]{freeman}
Freeman, K. 1970, ApJ, 160, 811

\bibitem[\protect\citename{Glicenstein }1999]{Gl99}
Glicenstein, J.F. 1999, A\&A, 343, 1025

\bibitem[\protect\citename{Graham }2002]{G02}
Graham, A.W. 2002, MNRAS, 334, 721

\bibitem[\protect\citename{Kerins et al. }2002]{KBS02}
Kerins, E., Binney, S., Silk, J. 2002, MNRAS, 332, L29

\bibitem[\protect\citename{Kravtsov et al. }1998]{Ketal98}
Kravtsov, A.V., Klypin, A.A., Bullock, J.S., Primack, J.R. 1998, ApJ, 502, 48

\bibitem[\protect\citename{Lasserre et al. }2000]{Letal00}
Lasserre, T. et al. 2000, A\&A, 355, L39

\bibitem[\protect\citename{Maller et al. }1997]{MFP97}
Maller, A.H., Flores, R.A., Primack, J.R. 1997, ApJ 486, 681

\bibitem[\protect\citename{Milgrom }1983]{M83}
Milgrom, M. 1983, ApJ, 270, 365

\bibitem[\protect\citename{Mo et al. }1998]{MMW98}
Mo, H.J., Mao, S., White, S.D.M. 1998, MNRAS, 295, 319

\bibitem[\protect\citename{Navarro, Frenck and White }1997]{NFW97}
Navarro, J.F., Frenck, C.S., White, S.D.M. 1997, ApJ, 490, 493

\bibitem[\protect\citename{Paczy\'nski }1986]{Pac86}
Paczy\'nski, B. 1986, ApJ, 304, 1

\bibitem[\protect\citename{Persic et al. }1996a]{PSS}
Persic, M., Salucci, P., Stel F. 1996a, MNRAS, 281, 27

\bibitem[\protect\citename{Persic et al. }1996b]{PSS2}
Persic, M., Salucci, P., Stel F. 1996b, MNRAS, 283, 1102

\bibitem[\protect\citename{Pfenniger et al. }1994]{PCM94}
Pfenniger, D., Combes, F., Martinet, L. 1994, A\&A, 285, 79

\bibitem[\protect\citename{Refsdal }1964]{Ref64}
Refsdal, S. 1964, MNRAS, 128, 307

\bibitem[\protect\citename{Salucci \& Borriello }2002]{SB02}
Salucci, P., Borriello, A. 2002, astro\,-\,ph/0203457

\bibitem[\protect\citename{Salucci \& Burkert }2000]{SBu00}
Salucci, P., Burkert, A. 2000, ApJ, 537, L9

\bibitem[\protect\citename{Schneider et al. }1992]{SEF}
Schneider, P., Ehlers, J., Falco, E.E. 1992, {\it Gravitational lenses}, Springer\,-\,Verlag, Berlin

\bibitem[\protect\citename{Sereno }2002]{S02}
Sereno, M. 2002, Phys. Lett. A, 305, 7

\bibitem[\protect\citename{Sereno }2003]{S03}
Sereno, M. 2003, Phys. Rev. D, 67, 064007

\bibitem[\protect\citename{Sereno \& Cardone }2002]{SCnoi02}
Sereno, M., Cardone, V.F. 2002, A\&A, 396, 393

\bibitem[\protect\citename{Sofue \& Rubin }2001]{SR01}
Sofue, Y., Rubin, V. 2001, ARA\&A, 39, 137

\bibitem[\protect\citename{Udalski et al. }1994]{ogle}
Udalski, A. et al. 1994, Acta Astron., 44, 165

\bibitem[\protect\citename{Vitvitska et al. }2002]{Vit02}
Vitvitska, M., Klypin, A., Kravtsov, A.V., Bullock, J.S., Wechsler, R.H., Primack, J.R. 2002, ApJ, 581, 799

\end{thebibliography}
\end{document}